\documentclass[aps,prd,reprint,twocolumn,superscriptaddress,longbibliography,nofootinbib,floatfix,showpacs]{revtex4-1}
\usepackage{epsfig}
\usepackage{amsmath,amssymb,amsfonts}
\usepackage{hyperref}
\usepackage{mathrsfs}
\usepackage{bbm}
\usepackage{slashed}
\usepackage{graphicx}
\usepackage{verbatim}

\usepackage{bm}

\usepackage[usenames]{color}

\usepackage{tikz}
\usetikzlibrary{calc,decorations.markings,decorations.pathmorphing,positioning,shapes,snakes,arrows}

\definecolor{darkgreen}{rgb}{0.2,0.6,0}

\newcommand{\be}{\begin{equation}}
\newcommand{\ee}{\end{equation}}
\newcommand{\bw}{\begin{widetext}}
\newcommand{\ew}{\end{widetext}}
\newcommand{\bi}{\begin{itemize}}
\newcommand{\ei}{\end{itemize}}
\newcommand{\bea}{\begin{eqnarray}}
\newcommand{\eea}{\end{eqnarray}}
\newcommand{\ud}{\mathrm{d}}

\newcommand{\LCm}{{\scriptscriptstyle -}} 

\usepackage[T1]{fontenc} \usepackage[latin1]{inputenc}

\begin{document}

\title{Resummation of quantum radiation reaction and induced polarization}

\author{Greger Torgrimsson}
\email{g.torgrimsson@hzdr.de}
\affiliation{Helmholtz-Zentrum Dresden-Rossendorf, Bautzner Landstra{\ss}e 400, 01328 Dresden, Germany}

\begin{abstract}

In a previous paper we proposed a new method based on resummations for studying radiation reaction of an electron in a plane-wave electromagnetic field. In this paper we use this method to study the electron momentum expectation value for a circularly polarized monochromatic field with $a_0=1$, for which standard locally-constant-field methods cannot be used. We also find that radiation reaction has a significant effect on the induced polarization, as compared to the results without radiation reaction, i.e. the Sokolov-Ternov formula for a constant field, or the zero result for a circularly monochromatic field. We also study the Abraham-Lorentz-Dirac equation using Borel-Pad\'e resummations. 
	
\end{abstract}	

\maketitle

\section{Introduction}

Radiation reaction (RR)~\cite{Burton:2014wsa,Blackburn:2019rfv}, i.e. the difference between the actual trajectory of a charge and the one predicted by the Lorentz force equation (with only the background field, no self-field), is becoming an important effect in high-intensity laser experiments~\cite{Cole:2017zca,Poder:2018ifi} (see also~\cite{Wistisen:2017pgr}). RR is an old topic which continues to be the subject of many theoretical studies~\cite{Ilderton:2013dba,Heinzl:2021mji,Ekman:2021vwg}.
Even in classical electrodynamics there continues to be a debate on which equation is correct. Perhaps more relevant for upcoming laser experiments is how to obtain quantum RR with spin etc. and e.g. how to go beyond the standard locally-constant-field (LCF) approximation. 
In~\cite{Torgrimsson:2021wcj} we developed a new method for calculating the momentum expectation value of an electron in a plane-wave background field. It is fully quantum and includes both real photon emissions and loops. This method is based on the use of first-order Mueller matrices as building blocks of higher-order probabilities~\cite{Dinu:2019pau,Torgrimsson:2020gws}.
In~\cite{Torgrimsson:2021wcj} we focused for simplicity on a constant field, but this method is not restricted to LCF or $a_0\gg1$. It works as long as the field is sufficiently long. In this paper we will consider both a constant field as well as a circularly polarized monochromatic field with $a_0=1$. The circular case provides a contrasting example compared to the constant field case, which we will now explain.

If an initially unpolarized electron is in the background field for a sufficiently long time, then it can become polarized. If one neglects RR and considers a constant field, the polarization is given by the famous Sokolov-Ternov result. This induced polarization is along the magnetic field direction. The Sokolov-Ternov effect was originally considered for electrons in storage-rings. However, one has a similar effect in other fields as well~\cite{DelSorbo:2017fod,Seipt:2019ddd,Li:2018fcz}, in particular plane waves. However, if one just lets an electron stay in the laser field then RR will eventually lead to a considerably change in the momentum. This is not included in the Sokolov-Ternov result. The method in~\cite{Torgrimsson:2021wcj} is not restricted to only the momentum expectation value; here we will use these methods to study how RR affects the induced polarization. 

For a linearly polarized field and in the LCF regime, it has been shown in~\cite{Chen:2019vly,Seipt:2019ddd} that the induced polarization tends to average out due to the fact that the magnetic field changes direction every half cycle. There are ways~\cite{Chen:2019vly,Seipt:2019ddd} to compensate for this by choosing some asymmetric field. One might wonder, though, if a circularly polarized field could be used instead to generate polarized particles. If one only considers the contribution from real photon emission then one would conclude that a circularly polarized field would be able to make the particles polarized along the propagation direction of the laser. However, at first order in $\alpha$, this is exactly canceled by the the cross term between the zeroth order amplitude and the first-order loop~\cite{KotkinUseLoops,Kotkin:1997me,Torgrimsson:2020gws} (see though~\cite{Karlovets:2011yy} for dependence on collision angle). Hence, if one neglects RR then there is no induced polarization for a circularly polarized field. Here we will use the methods of~\cite{Torgrimsson:2021wcj} to show that there is in fact a nonzero induced polarization if one takes RR into account.

Polarization effects of particles in a circularly polarized field in LCF have been studied in~\cite{Li:2020bwo,Li:2019oxr}. There is a crucial difference here. For a circular field in LCF, one needs to start with polarized particles in the initial state to have longitudinally polarized particles in the final state. Here we are instead interested in a circular field with $a_0\sim1$ in which, as we will show, even an initially unpolarized particle may become longitudinally polarized. In this regime, spin up and down the longitudinal direction plays essentially the same role as spin up and down the magnetic field direction for a linearly polarized field.  

This paper is organized as follows. In Sec.~\ref{RR and the Sokolov-Ternov effect} we study how RR affects the Sokolov-Ternov polarization for a constant field. In Sec.~\ref{Circularly polarized field} we show the details of how to use the general methods in~\cite{Torgrimsson:2021wcj} for a circularly polarized monochromatic field with $a_0=1$ (i.e. beyond LCF), and we apply these methods to both the momentum expectation value and the induced polarization. In order to gain some insights into how the approximation in~\cite{Torgrimsson:2021wcj} relates to the full QED result (which cannot be calculated), in Sec.~\ref{Resummation of LAD} we compare its leading classical limit, which we in~\cite{Torgrimsson:2021wcj} showed is exactly equation to the LL equation, with a resummation of LAD.

\section{RR and the Sokolov-Ternov effect}\label{RR and the Sokolov-Ternov effect}

In~\cite{Torgrimsson:2021wcj} we showed that for a constant field we have
\be\label{recursionLCF}
{\bf M}^{(n)}=\int_0^1\frac{\ud q}{n\chi}({\bf M}^{\rm C}\cdot{\bf M}^{(n-1)}(\chi(1-q))+{\bf M}^{\rm L}\cdot{\bf M}^{(n-1)}(\chi)) \;,
\ee
where the $\mathcal{O}(\alpha)$ building blocks are expressed in terms of Airy functions and $\text{Ai}_1(\xi)=\int_\xi^\infty\ud x\,\text{Ai}(x)$, and if we restrict to an initial electron that is either unpolarized or polarized up or down the magnetic field direction then the $4\times4$ Mueller matrices reduce to the following $2\times2$ matrices
\be
{\bf M}^{\rm C}=\begin{pmatrix}-\text{Ai}_1(\xi)-\kappa\frac{\text{Ai}'(\xi)}{\xi} & \frac{q}{s_1}\frac{\text{Ai}(\xi)}{\sqrt{\xi}} \\ q\frac{\text{Ai}(\xi)}{\sqrt{\xi}} & -\text{Ai}_1(\xi)-2\frac{\text{Ai}'(\xi)}{\xi}\end{pmatrix} \;,
\ee
\be
{\bf M}^{\rm L}=\begin{pmatrix}\text{Ai}_1(\xi)+\kappa\frac{\text{Ai}'(\xi)}{\xi} & -q\frac{\text{Ai}(\xi)}{\sqrt{\xi}} \\ -q\frac{\text{Ai}(\xi)}{\sqrt{\xi}} & \text{Ai}_1(\xi)+\kappa\frac{\text{Ai}'(\xi)}{\xi}\end{pmatrix} \;,
\ee
$\xi=(r/\chi)^{2/3}$, $r=(1/s_1)-1$, $\kappa=(1/s_1)+s_1$ and $s_1=1-q$. In~\cite{Torgrimsson:2021wcj} we considered the expectation value of the longitudinal momentum $\langle kP\rangle$ and then the first term is given by ${\bf M}^{(0)}=b_0{\bf 1}$. In this section we will consider the probability of transition between Stokes vector ${\bf N}_i$ and ${\bf N}_f$, for which we have $\mathbb{P}=(1/2){\bf N}_i\cdot{\bf M}\cdot{\bf N}_f$, where ${\bf M}=\sum_{n=0}^\infty T^n{\bf M}^{(n)}$ and ${\bf M}^{(n)}$ is obtained from the same~\eqref{recursionLCF} but with ${\bf M}^{(0)}={\bf 1}$. 
The effective expansion parameter is 
\be\label{TLCF}
\text{constant field:}\qquad
T=\alpha a_0\Delta\phi 
\ee
where $\Delta\phi$ is the length of the field. 

We can use~\eqref{recursionLCF} to first calculate the orders separately and then resum the $\alpha$ expansion afterwards. Or we can resum~\eqref{recursionLCF} right from the start, which gives ${\bf M}$ as the solution to an integro-differential equation 
\be\label{integroDiffLCF}
\frac{\partial{\bf M}}{\partial T}=\int_0^1\frac{\ud q}{\chi}({\bf M}^{\rm C}\cdot{\bf M}(\chi[1-q])+{\bf M}^{\rm L}\cdot{\bf M}(\chi)) \;,
\ee
where ${\bf M}(T=0)={\bf M}^{(0)}$.

\subsection{Standard Sokolov-Ternov, no RR}

If we neglect RR then we should obtain the standard Sokolov-Ternov result. In our approach, we turn off RR by replacing $\chi(1-q)\to\chi$ in the first term in~\eqref{recursionLCF}. 
The first-order Mueller matrix is given by
\be
{\bf M}_1=\begin{pmatrix}0&\chi^2\\0&-\frac{5\sqrt{3}}{8}\chi^2\end{pmatrix}
\ee
and with no RR it is straightforward to sum up all orders, which gives
\be\label{STallOrder}
\mathbb{P}=\frac{1}{2}{\bf N}_0\cdot\begin{pmatrix}1&\frac{8}{5\sqrt{3}}\left(1-\exp\left[-\frac{5\sqrt{3}}{8}\chi^2T\right]\right)\\0&\exp\left[-\frac{5\sqrt{3}}{8}\chi^2T\right]\end{pmatrix}\cdot{\bf N}_f \;,
\ee
which agrees with the results in~\cite{Sokolov:1963zn,BaierSokolovTernov}. For a very long pulse, $T\gg1/\chi^2$ and with an initially unpolarized state, 
\be\label{STallOrderAsymptotic}
\mathbb{P}=\frac{1}{2}\left\{1,\frac{8}{5\sqrt{3}}\right\}\cdot{\bf N}_f \;,
\ee
so the degree of the induced polarization is $8/(5\sqrt{3})\approx0.92$. Note that this is just a constant, in particular, independent of the momentum $\chi$. However, the time scale for this is $T\sim1/\chi^2$, which is slower than the time scale for RR, which causes the longitudinal momentum to decrease as $\chi(T)=\chi/(1+[2/3]\chi T)$. Since for large $T$ the combination $\chi^2(T)T\sim1/T$ is actually small rather than large, this suggests that the result with RR included will be very different from~\eqref{STallOrder}. 

\subsection{RR included, low energy}\label{RR included, low energy}

To see the effects of RR, we consider first the leading order in $\chi\ll1$. Inserting the ansatz
\be\label{MnAB}
{\bf M}_n=\chi^{n+1}\begin{pmatrix}0&B_n\\0&A_n\end{pmatrix}
\ee
for $n\geq1$ into~\eqref{recursionLCF},
we find $B_n=-(2/3)B_{n-1}$ and $A_n=-(2/3)A_{n-1}$. Thus we have a simple geometric series and we find
\be
\mathbb{P}=\frac{1}{2}{\bf N}_0\cdot\left[{\bf 1}+\frac{T\chi^2}{1+\frac{2}{3}T\chi}\begin{pmatrix}0&1\\0&-\frac{5\sqrt{3}}{8}\end{pmatrix}\right]\cdot{\bf N}_f \;.
\ee
Note that this varies on the same time scale as the momentum expectation value, i.e. $T\sim1/\chi$. For an asymptotically long pulse, $T\gg1/\chi$, and for an initially unpolarized electron we have
\be\label{STwithRRasymptotic}
\mathbb{P}=\frac{1}{2}\left\{1,\frac{3\chi}{2}\right\}\cdot{\bf N}_f \;.
\ee
This is quite different from the standard Sokolov-Ternov result~\eqref{STallOrderAsymptotic}: The asymptotic result is reached much faster ($T\gg1/\chi$), but is a much lower degree of polarization for $\chi\ll1$. However, \eqref{STwithRRasymptotic} suggests that we might be able to compensate for the negative effect of RR by increasing $\chi$. Since the degree of polarization cannot be larger than $1$, the leading order~\eqref{STwithRRasymptotic} cannot be used for larger $\chi$. So, we now turn to a numerical treatment.

\subsection{RR included, higher energy}

\begin{figure}
\includegraphics[width=\linewidth]{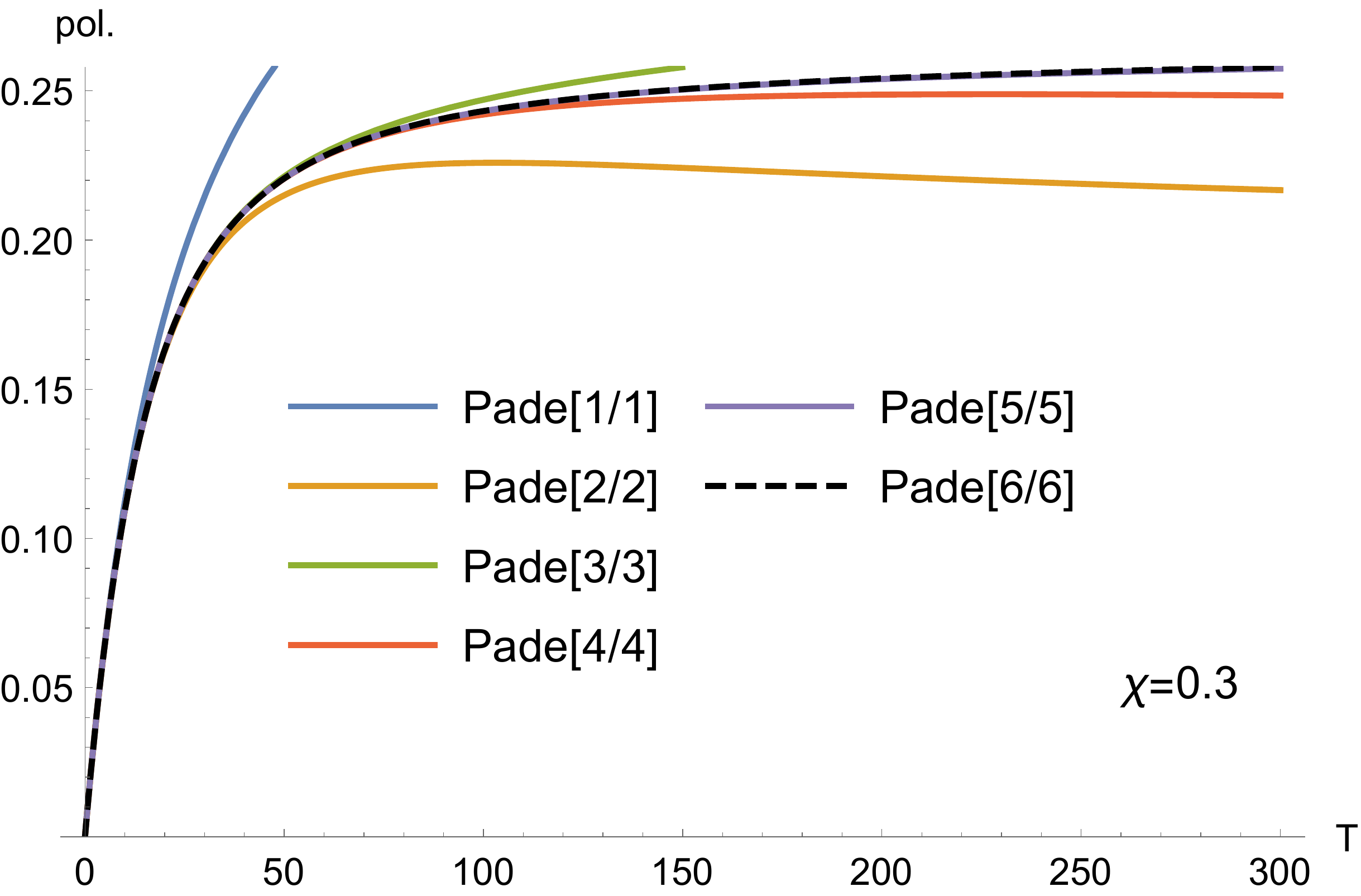}
\caption{The polarization induced on an initially unpolarized electron. Pade$[n/n]$ refers to diagonal Pad\'e approximants $(\sum_{i=1}^n c_i T^i)/(1+\sum_{i=1}^n d_i T^i)$. The leading order~\eqref{STwithRRasymptotic} leads to a large overestimation for this value of $\chi$ and is therefore not shown.}
\label{STRRchi03}
\end{figure}

The orders ${\bf M}_n$ can be calculated separately either by resummation of the $\chi$ expansion or with numerical interpolation functions exactly in the same way as in~\cite{Torgrimsson:2021wcj}. The only difference is that in~\cite{Torgrimsson:2021wcj} we considered the expectation value of the  longitudinal momentum, with ${\bf M}_0=b_0{\bf 1}$, while now we have ${\bf N}_f=\{0,1\}$. 
However, in~\cite{Torgrimsson:2021wcj} we used the ``standard'' Borel transform to resum the $\chi$ expansions, while here we are going to use a slight modification.

We begin by noting that the first order can be expressed as a Mellin integral 
\be\label{M1Mellin}
{\bf M}_1=\int\frac{\ud S}{2\pi i}\frac{\pi}{4}\frac{3^{S/2}}{\chi^S}\frac{S(1+S)}{\sin[\pi S]}\begin{pmatrix}0&\frac{\Gamma\left[1-\frac{3S}{2}\right]}{\Gamma\left[1-\frac{S}{2}\right]} \\0&-\frac{\sqrt{3}\Gamma\left[\frac{1}{2}-\frac{3S}{2}\right]}{\Gamma\left[\frac{1}{2}-\frac{S}{2}\right]}\end{pmatrix} \;,
\ee
where the integration contour goes through the real axis at $-2<\text{Re }S<1/3$. The $\chi<1$ expansion is obtained directly from the residues of the poles at $S=-n$, $n=2,3,...$. For large $n$ we have for the coefficient of $\chi^m$
\be\label{M1largeOrder}
{\bf M}_1^{(m)}\sim\frac{1}{4}\sqrt{\frac{3}{2\pi}}\left(-\frac{3}{2}\right)^m\Gamma\left[m+\frac{5}{2}\right]\begin{pmatrix}0&1+\mathcal{O}(1/m)\\0&-1+\mathcal{O}(1/m)\end{pmatrix} \;.
\ee 
Of course, for this order we already have an exact expression, e.g.~\eqref{M1Mellin}, but~\eqref{M1largeOrder} will help us to guess a convenient resummation for ${\bf M}_n$, $n>1$. We write ${\bf M}_n$ as in~\eqref{MnAB} but where $A_n$ and $B_n$ are now functions of $\chi$. We focus on $B_n$. Its expansion $B_n=\sum_{m=0}^\infty B_n^{(m)}\chi^m$ starts with the constants $B_n^{(0)}$ obtained in Sec.~\ref{RR included, low energy}. In contrast to the $n=1$ case, for $n>1$ we only have the first $\sim100$ terms in $\chi$. From~\eqref{M1largeOrder} we take 
\be
c^m\Gamma\left[m+b_n\right] \;,
\ee 
as an ansatz for the large-order scaling of $B_n^{(m)}$. We take $c=3/2$ and determine $b$ numerically. We find $b_n=3+(3/2)n$ for $n\geq1$. Note that the value of $b_n$ does not need to be perfect (in fact, $b_n=3+(3/2)n$ might be only an approximation), but we see that at larger $n$ $b_n$ is considerably larger than $b_1=9/2$, which one might expect to be significant. Thus, instead of the standard Borel transform, we use the Borel-Leroy transform (the factor of $c^m$ is just a variable rescaling)
\be\label{BorelLoreyTransform}
BL[B_n](t)=\sum_{m=0}^\infty\frac{B_n^{(m)}}{c^m\Gamma\left[m+b_n\right]}t^m \;.
\ee
This can now be resummed with a Pad\'e approximant. We have used\footnote{In some cases one may have to choose slightly different Pad\'e orders due to the appearance of spurious poles.} $[40/(40+\text{ceil}[b_n]+1)]$. The final result is obtained from a Laplace integral
\be
B_n^{\rm(resum)}(\chi)=\int_0^\infty\frac{\ud t}{x}\left(\frac{t}{x}\right)^{b_n-1}e^{-t/x}PBL[B_n](t) \;,
\ee
where $x=c\chi$. Now we have the first $\sim10$ to $20$ terms in the $\alpha$ expansion.

The next step is to resum the $\alpha$ expansion.
To have a nonzero (and finite) limit as $T\to\infty$ we resum this expansion with diagonal Pad\'e approximants, $[n/n]$. The result is shown in Fig.~\ref{STRRchi03}. As in~\cite{Torgrimsson:2021wcj}, we again find a fast convergence of the Pad\'e approximants. For this particular value of $\chi$, we are close to the asymptotic limit at $T=100$ and for $T<100$ we see that the resummation has already converged to a good precision with Pad\'e$[4/4]$, which is obtained with the first eight terms in the $\alpha$ expansion. The leading order~\eqref{STwithRRasymptotic} gives a rather large overestimation at $\chi=0.3$, but Fig.~\ref{STRRchi03} shows that the exact result still gives a significant degree of polarization.

\section{Circularly polarized field}\label{Circularly polarized field}

In~\cite{Torgrimsson:2021wcj} we considered a constant field. Here we will consider a circularly polarized monochromatic field. As a constant field can be seen as a first step towards LCF, a monochromatic field can be seen as the first step towards LMF. While approaches based on LCF only work for large $a_0$, LMF works also for smaller $a_0$ provided the pulse length is sufficiently long. In this section we will therefore focus on $a_0\sim1$. The effective expansion parameter is
\be\label{TLMF}
\text{circular field:}\qquad
T=\alpha\Delta\phi \;,
\ee
for ${\bf a}(\phi)=a_0\{\sin\phi,\cos\phi,0\}$ and $0<\phi<\Delta\phi$.

\subsection{Definitions}

In general the Mueller matrices are $4\times4$. For a constant field one can reduce this to $2\times2$ if one considers initial and final electrons that are either unpolarized or polarized parallel or anti-parallel to the magnetic field. For a circularly polarized field (with $a_0\sim1$) the general $4\times4$ matrices can be found in~\cite{Torgrimsson:2020gws}. However, also in this case
one can reduce the Mueller matrices to $2\times2$, but in this case the special spin direction is parallel/anti-parallel to the wave vector of the field. 
The recursive and integro-differential equations for the circular field are obtained from~\eqref{recursionLCF} and~\eqref{integroDiffLCF} by replacing $\chi\to b_0$ and, for initial and final states that are unpolarized or polarized along the wave vector of the field, the first-order Mueller-matrix building blocks are given by~\cite{Torgrimsson:2020gws}
\be
{\bf M}^{\rm C}=\begin{pmatrix} \frac{\kappa}{2}\mathcal{J}_1-\mathcal{J}_0 & \frac{q}{2} \left(1+\frac{1}{s_1}\right)\mathcal{J}_2 \\ \frac{q}{2} \left(1+\frac{1}{s_1}\right)\mathcal{J}_2 &-\frac{q^2}{2s_1}\mathcal{J}_0+\frac{\kappa}{2}(\mathcal{J}_1-\mathcal{J}_0) \end{pmatrix}
\ee
and
\be
{\bf M}^{\rm L}=\begin{pmatrix} -\frac{\kappa}{2}\mathcal{J}_1+\mathcal{J}_0 & -\frac{q}{2} \left(1+\frac{1}{s_1}\right)\mathcal{J}_2 \\ -\frac{q}{2} \left(1+\frac{1}{s_1}\right)\mathcal{J}_2 & -\frac{\kappa}{2}\mathcal{J}_1+\mathcal{J}_0 \end{pmatrix}
\ee
where $s_1=1-q$, $\kappa=(1/s_1)+s_1$, $r=(1/s_1)-1$,
\be
\mathcal{J}_0=\frac{i}{2\pi}\int\frac{\ud\theta}{\theta}\exp\left\{\frac{ir}{2b_0}\Theta\right\}=\sum_n J_n^2(z) \;,
\ee
\be
\begin{split}
\mathcal{J}_1=&-2a_0^2(u)\frac{i}{2\pi}\int\frac{\ud\theta}{\theta}\sin^2\left(\frac{\theta}{2}\right)\exp\left\{\frac{ir}{2b_0}\Theta\right\} \\
=&\frac{a_0^2(u)}{2}\sum_n[J_{n+1}^2(z)+J_{n-1}^2(z)-2J_n^2(z)]
\end{split}
\ee
and
\be
\begin{split}
\mathcal{J}_2=&-\frac{a_0^2(u)}{2\pi}\int\frac{\ud\theta}{\theta}\left(\sin\theta-\frac{4}{\theta}\sin^2\frac{\theta}{2}\right)\exp\left\{\frac{ir}{2b_0}\Theta\right\} \\
=&\frac{1}{2}\sum_n\left(\frac{b_0z^2}{nr}-a_0^2(u)\right)[J_{n-1}^2(z)-J_{n+1}^2(z)] \;,
\end{split}
\ee
where the first version of these functions involve the effective mass
\be\label{MLMF}
\Theta=\theta M^2=\theta\left[1+a_0^2(u)\left(1-\text{sinc}^2\frac{\theta}{2}\right)\right] \;,
\ee
$a_0(u)=a_0h(u)$ and $\text{sinc}(x)=\sin(x)/x$ and the singularity at $\theta=0$ is avoided using a contour equivalent to $\theta\to\theta+i\epsilon$, and the second version is expressed in terms of Bessel functions $J_n$ with argument
\be
z=a_0(u)\frac{r}{b_0}\left[\frac{2b_0n}{r}-1-a_0^2(u)\right]^\frac{1}{2} \;,
\ee
and the sum over $n$ is restricted by
\be
n\geq\frac{r}{2b_0}[1+a_0^2(u)] \;.
\ee
In this paper we will for simplicity consider a monochromatic field, i.e. $a_0(u)=a_0=\text{constant}$. We change integration variable from $q$ to $\gamma=r/b_0$.

\subsection{Integrals for the $b_0$ expansion}

To obtain the $b_0$ expansions we need the following integrals
\be
\mathcal{I}_i^{(n)}=\int_0^\infty\ud\gamma\;\gamma^n\mathcal{J}_i(\gamma) \;.
\ee
We have calculated these using the $\theta$-integral rather than the Bessel-function representations. We have
\be
\mathcal{I}_0^{(n)}=\int\ud\theta\frac{i2^nn!}{\pi\theta(-i\Theta)^{1+n}} \;,
\ee
and similarly for $\mathcal{I}_1^{(n)}$ and $\mathcal{I}_2^{(n)}$. With a partial integration one can show that $\mathcal{I}_2^{(n)}=-n\mathcal{I}_0^{(n-1)}$, so we focus on $\mathcal{I}_0^{(n)}$ and $\mathcal{I}_1^{(n)}$. The factorial growth of these integrals is responsible for the factorial growth of the $b_0$ expansion. We can obtain a large order expansion by exponentiating
\be
\frac{1}{(-i\Theta)^n}=\exp\left(-n\ln[-i\Theta]\right)
\ee
and then performing the $\theta$ integral with the saddle-point method. The saddle-point, $\theta_s$, is determined by a transcendental equation which can easily be solved numerically. Then one can change integration variable from $\theta=\theta_s+\delta\theta/\sqrt{n}$ to $\delta\theta$, expand the integrand in $1/n$ and perform the resulting Gaussian integrals (analytically). We find
\be\label{I0intLargen}
\mathcal{I}_0^{(n)}\approx(1.588...)^n\Gamma\left[n+\frac{1}{2}\right](0.34...-\frac{0.21...}{n}+\frac{0.22...}{n^2}-...)
\ee
and
\be
\mathcal{I}_1^{(n)}\approx(1.588...)^n\Gamma\left[n+\frac{1}{2}\right](0.74...+\frac{0.017...}{n}-\frac{0.019...}{n^2}-...)
\ee
where the coefficients have been obtained to high precision and have not been rounded for the digits shown above. A direct sum of the first 11 terms, i.e. including the term with $1/n^{10}$, has a precision of $\sim10^{-11}$ for $n=10$ and $\sim10^{-22}$ for $n=100$. 

We still need to obtain $\mathcal{I}_i^{(n)}$ for finite $n$. We have found that in the process of calculating a large number of terms in the $b_0$ expansion and several terms in the $\alpha$ expansion, and then resumming these expansions, the precision can be reduced by several orders of magnitude. However, this is not a problem, because we can obtain a very high precision of $\mathcal{I}_i^{(n)}$. For odd $n$ the integrals are simply given by $-i\pi$ times the residue at $\theta=0$, which gives rational numbers. So, for odd $n$ we have exact expressions. For even $n$ we could in principle close the contour in the upper half plane and use the residue theorem for the poles there. However, there are an infinite number of poles, and, although each pole can be obtained quickly and to a high precision using 
\be
2\pi\left(n+\frac{1}{2}\right)+i\ln\left[2(2\pi)^2\left(n+\frac{1}{2}\right)^2\right]
\ee
as starting point for a numerical root finding, the sum over these poles converges slowly.

It is actually faster to just perform the integrals directly by integrating along the real axis except for a detour above $\theta=0$ at a distance $|\theta|\sim1$. With this approach we have obtained the integrals to a precision of $\sim10^{-40}$ for all the relevant $n$. However, the first couple of $n$ takes a rather long time to compute. For larger $n$, on the other hand, the integrals converge very quickly, so for $n\gtrsim20$ (the largest $n$, e.g. $n_{\rm max}=130$, needs to be about the same as the maximum order in $b_0$) we can quickly obtain a precision of e.g. $\sim10^{-100}$. 

We can also obtain such a high precision for lower $n$ with the following approach, where we use a third expansion. This is essentially a perturbative expansion in the amplitude of the field, $a_0$, except that, as seen from~\eqref{MLMF}, it is more natural to expand in $\delta:=a_0^2/(1+a_0^2)$ instead. Then the expansion of $1/(1-\delta\text{sinc}^2[\theta/2])^{1+n}$ is a binomial series. Each term in the integrand now has the form $(1/\theta^{j})\text{sinc}^{2m}[\theta/2]$. We use the binomial theorem to write this as a sum of terms with $e^{i(m-k)\theta}$, where $0\leq k\leq2m$. Now we can use the residue theorem to calculate each term. Only those terms with $m<k\leq2m$ contributes. We find
\be\label{I0mksum}
\begin{split}
\mathcal{I}_0^{(n)}=&(1-\delta)^{1+n}\sum_{m=1}^\infty\delta^m\sum_{k=m+1}^{2m} \\
\times&\frac{(-1)^k2^{1+n}(k-m)^{2m+n+1}(2m)!(m+n)!}{m!k!(2m+n+1)!(2m-k)!} 
\end{split}
\ee
and
\be\label{I1mksum}
\begin{split}
\mathcal{I}_1^{(n)}=&(1-\delta)^{n}\sum_{m=1}^\infty\delta^m\sum_{k=m+1}^{2m} \\
\times&\frac{(-1)^k2^n(k-m)^{2m+n-1}(2m)!(m+n-1)!}{(m-1)!k!(2m+n-1)!(2m-k)!} \;.
\end{split}
\ee
The sums over $m$ have a finite radius of convergence. To see this, we go back a couple of steps and write $\text{sinc}^{2m}[\theta/2]=\exp\{m\ln\left(\text{sinc}^m[\theta/2]\right)\}$, and then the large $m$ behavior can be obtained by rescaling $\theta\to\theta/\sqrt{m}$ and expanding the integrand in $1/m$. For even $n$ (recall that odd $n$ are anyway simple) we find
\be\label{largem0}
\mathcal{I}_0^{(n,m\gg1)}\approx\frac{(1-\delta)^{1+n}m^{(n+1)/2}(m+n)!}{2\times3^{(n+1)/2}m!\Gamma\left[\frac{3+n}{2}\right]}
\ee
and
\be\label{largem1}
\mathcal{I}_1^{(n,m\gg1)}\approx\frac{(1-\delta)^nm^{(n-1)/2}(m+n-1)!}{3^{(n-1)/2}(m-1)!\Gamma\left[\frac{1+n}{2}\right]}
\ee
or to leading order
$\mathcal{I}_0^{(n,m\gg1)}\sim\text{const.}\times m^{(3n+1)/2}$ and
$\mathcal{I}_1^{(n,m\gg1)}\sim\text{const.}\times m^{(3n-1)/2}$.
From $\mathcal{I}_i^{(n,m+1)}/\mathcal{I}_i^{(n,m)}\to1$ and the ratio test we find that the radius of convergence is $\delta=1$, which is fortunate since we always have $\delta:=a_0^2/(1+a_0^2)<1$. Hence, apart from the partial resummation achieved by using $\delta$ rather than $a_0$, no additional resummation is needed\footnote{If $\delta$ is close to $1$ it might be useful to find some resummation. However, then $a_0$ would have to be large, which might mean that a LCF expansion in $1/a_0$ would anyway be more convenient.} and~\eqref{I0mksum} can therefore be summed directly. \eqref{I0mksum} allows us to obtain a very high precision quickly. It is much faster than the direct numerical integration for lower $n$, but slower for large $n$. Thus, we have used~\eqref{I0mksum} for the first $\sim20$ terms and a direct integration for $n\gtrsim20$, in both cases obtaining a precision of $\lesssim10^{-100}$. Summing the first $1\leq m\leq500$ terms in~\eqref{I0mksum} is enough to achieve this precision for $n\sim20$. Fewer terms are needed for lower $n$, while more terms are needed for larger $n$. The precision can be checked either by summing many more terms and checking the relative precision, or by estimating the size of the remainder by summing the large order approximation~\eqref{largem0} and~\eqref{largem1} from e.g. $m=500$ to $\infty$ in terms of polylogarithms, or by comparing with the direct integration.

\subsection{Momentum expectation value}

\begin{figure}
\includegraphics[width=\linewidth]{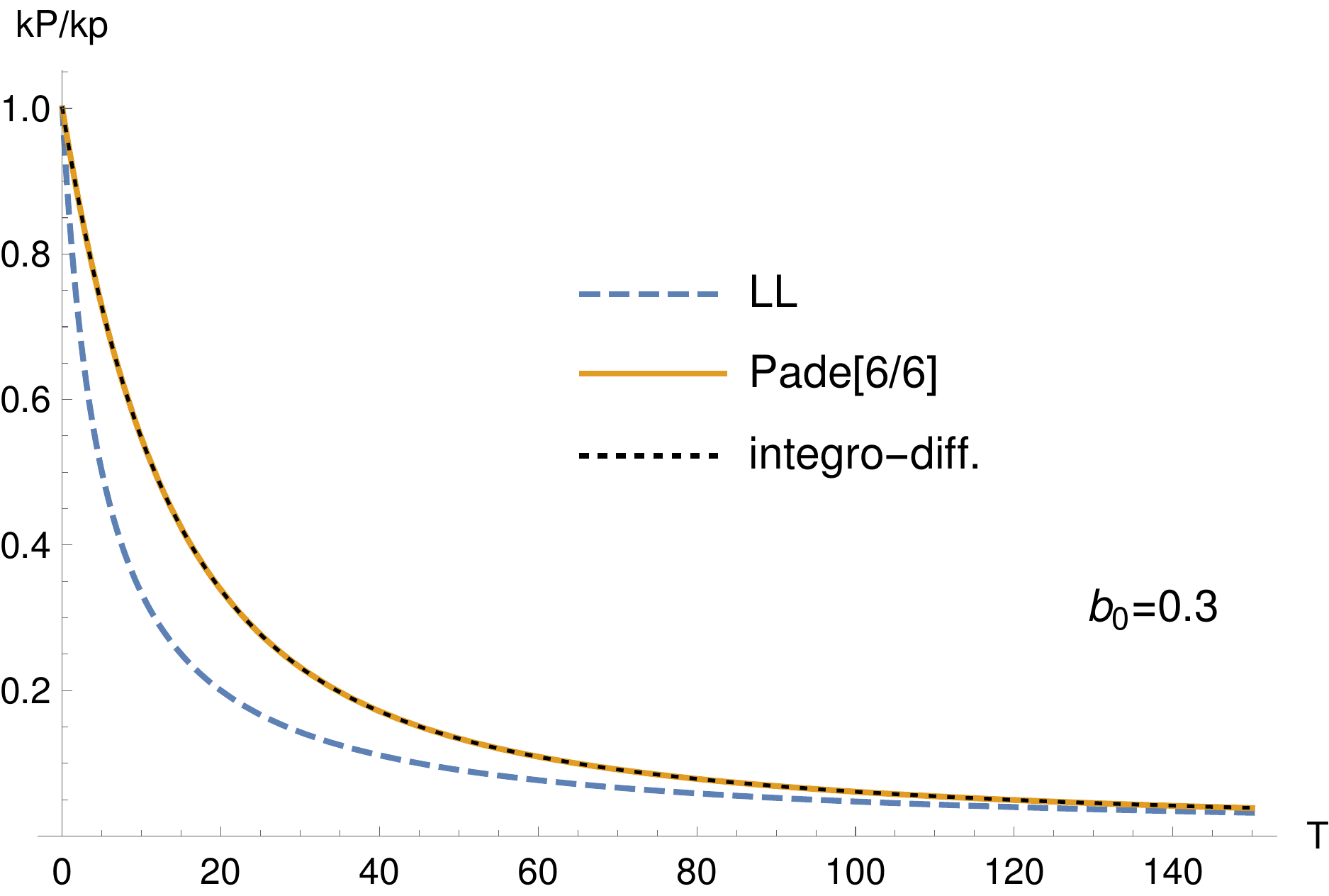}
\includegraphics[width=\linewidth]{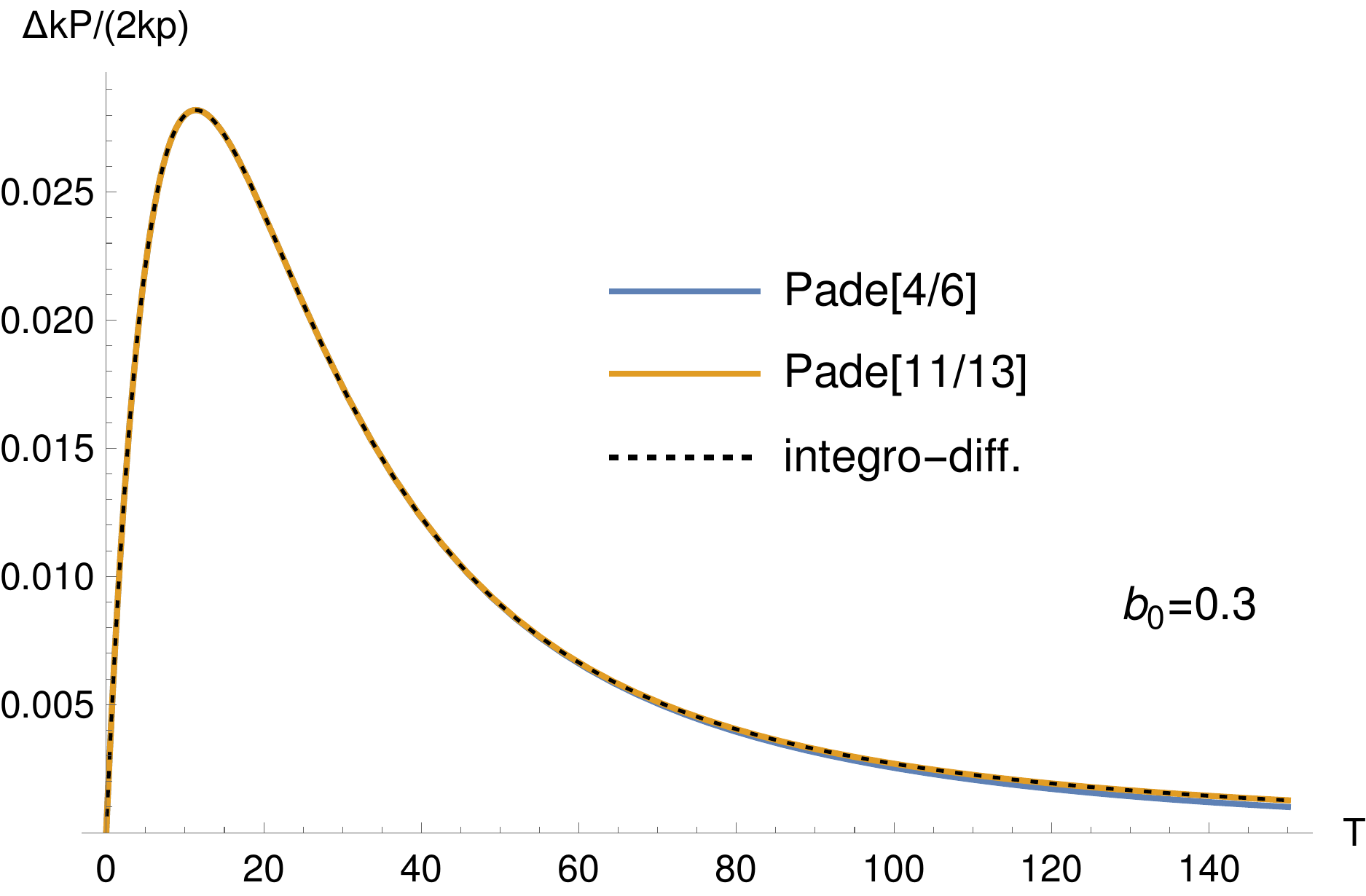}
\caption{Momentum expectation value for a circularly polarized monochromatic field.}
\label{RRcircFig}
\end{figure}

We consider first the momentum expectation value and to leading order in $\chi\ll1$. We use the ansatz ${\bf N}^{(n)}\approx a_0^{2n}b_0^{n+1}\{A_n+C_n b_0,B_n b_0\}$. We find a recursive formula, $A_n=-(2/3)A_{n-1}$ and
\be
B_n=\frac{2}{3}\left(A_{n-1}-\left[1+\frac{1}{n}\right]B_{n-1}\right)
\ee
so $C_n$ is not needed, with initial conditions $A_1=-2/3$ and $B_1=2/3$. With Eq.~(2.2.7) in~\cite{BenderOrszag} we can solve this in terms of the harmonic number $H_n$,
\be
B_n=\left(-\frac{2}{3}\right)^n(1+n)(1-H_n) \;.
\ee
Resumming this gives
\be
\sum_{n=1}^\infty\{0,1\}\cdot{\bf N}^{(n)}T^n=\frac{b_0^2\ln\left[1+\frac{2}{3}a_0^2b_0T\right]}{\left[1+\frac{2}{3}a_0^2b_0T\right]^2} \;.
\ee
Apart from some constants, this is essentially the same function\footnote{Note that $T$ includes a factor of $a_0$ in the constant field case~\eqref{TLCF}, but not in the circular case~\eqref{TLMF}.} as the spin difference for a constant field~\cite{Torgrimsson:2021wcj}, even though we are considering a different spin direction.
 
At higher energy we can use the same resummation methods or solve the integro-differential equation as in the constant-field case~\cite{Torgrimsson:2021wcj}. 
For the resummation approach we have used the Borel-Leroy method~\eqref{BorelLoreyTransform} with $c=1.588...$ (same as in~\eqref{I0intLargen}), $b_n=3+(3/2)n$ and $b_n=4+(3/2)n$ for $\langle kP\rangle(\uparrow)+\langle kP\rangle(\downarrow)$ and $\langle kP\rangle(\uparrow)-\langle kP\rangle(\downarrow)$, respectively, and $[40/(40+\text{ceil.}[b_n]+1)]$ or a nearby approximant if there is a spurious pole for some $n$. 

The integro-differential equation has been solved using the midpoint method, i.e. we make an interpolation function for the $b_0$ dependence of
\be
{\bf M}(T_{n+1/2})={\bf M}(T_n)+\frac{dT}{2}\frac{\partial{\bf M}(T_n)}{\partial T}
\ee
where the derivative is obtained from (the circular version of) \eqref{integroDiffLCF}, and then we use this to compute an interpolation function at the next step
\be
{\bf M}(T_{n+1})={\bf M}(T_n)+dT\frac{\partial{\bf M}(T_{n+1/2})}{\partial T} \;,
\ee
and so on.
In all cases in this paper we have used a step size $dT=1/10$. This integro-differential approach is much slower than the resummation approach.

Fig.~\ref{RRcircFig} shows that both approaches give the same result. These methods are the same as for the constant-field case. Fig.~\ref{RRcircFig} shows that also the results are very similar in shape as in the constant field case.

\subsection{Sokolov-Ternov effect}

\begin{figure}
\includegraphics[width=\linewidth]{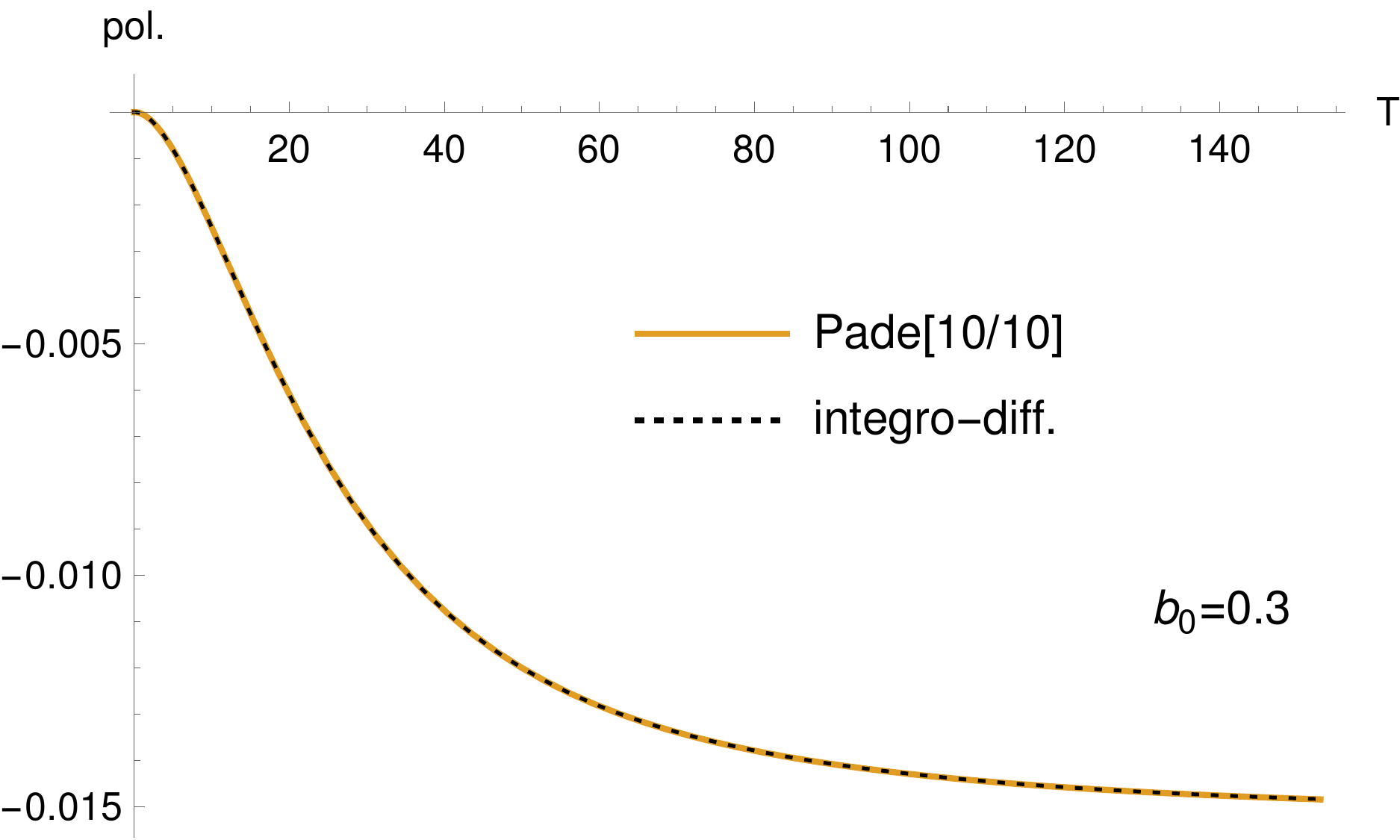}
\includegraphics[width=\linewidth]{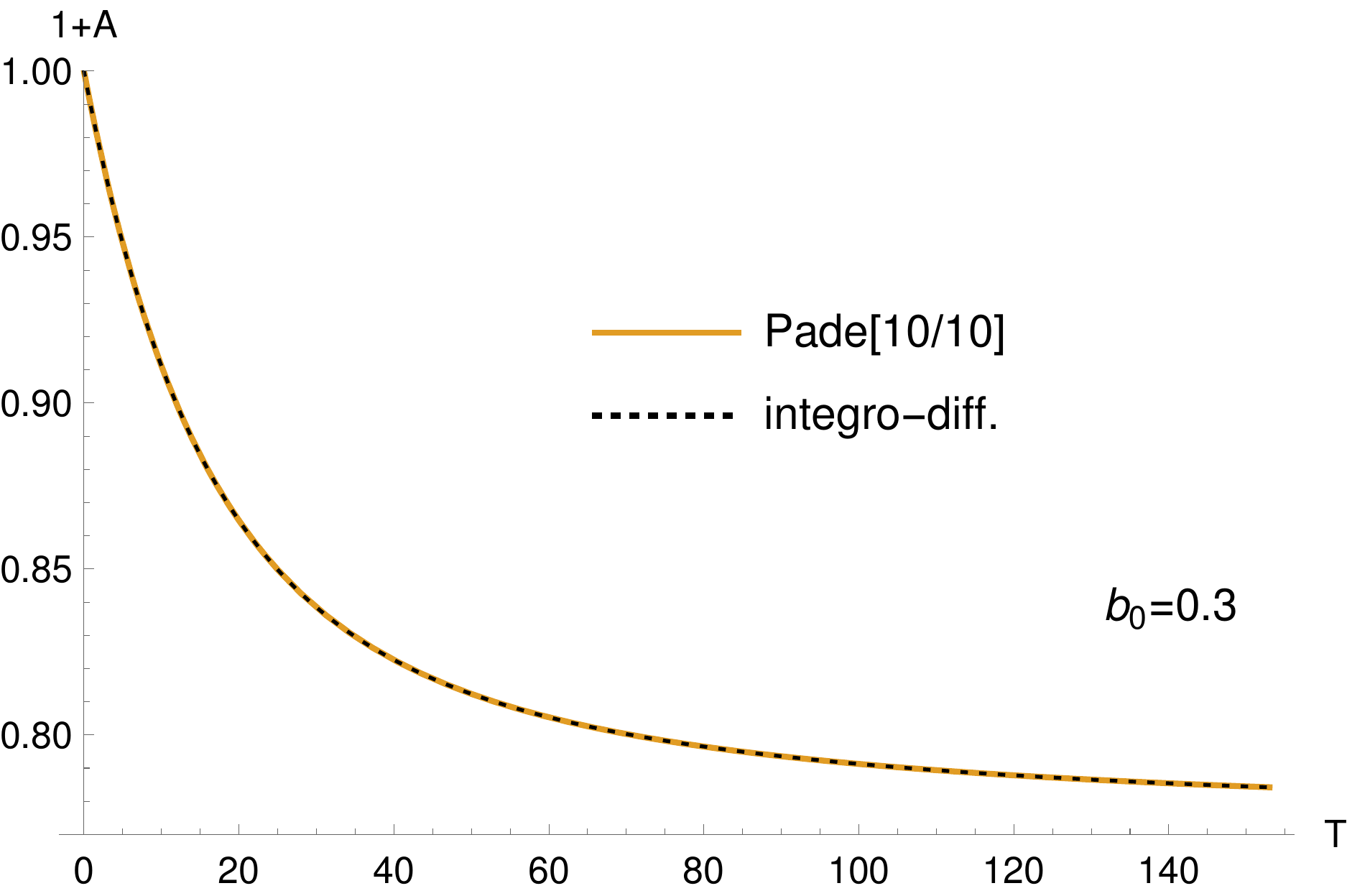}
\caption{$B$ and $1+A$ in~\eqref{ABmatrix} for a circularly polarized monochromatic field.}
\label{STRRcircFig}
\end{figure}

Without RR, three of the four elements in the sum ${\bf M}^{\rm C}+{\bf M}^{\rm L}\propto-\hat{\bf k}\hat{\bf k}$ cancel, and the sum of all orders gives
\be
\mathbb{P}=\frac{1}{2}{\bf N}_0\cdot\begin{pmatrix} 1&0\\0&\exp\left[-\mathcal{T}\hat{M}_1\right]\end{pmatrix}\cdot{\bf N}_f \;.
\ee
The factor $\hat{M}_1$ can be obtained directly by integrating ${\bf M}^{\rm C}+{\bf M}^{\rm L}$. But we do not need to obtain it in order to see that without RR the degree of polarization can only be reduced: If we start with an unpolarized electron ${\bf N}_0=\{1,0\}$ then there is no effect at all, while if we start with a polarized electron ${\bf N}_0=\{1,\pm1\}$ then the degree of polarization decreases as the pulse length $\mathcal{T}$ increases. We will now show that this changes if RR is included.

To see what happens when RR is included, we consider first the low-energy limit. With the ansatz
\be
{\bf M}_n=a_0^{2(n-1)}b_0^{n+1}\begin{pmatrix}0&B_nb_0\\0&A_n+C_n b_0\end{pmatrix}
\ee  
for $n\geq1$,
we find $A_n=(-2/3)^{n-1}A_1$ and
\be
B_n=-\frac{2}{3}\left(1+\frac{1}{n}\right)B_{n-1}-\left(-\frac{2}{3}\right)^{n-1}A_1 \;,
\ee
while $C_n$ is not needed. We can solve this using again Eq.~(2.2.7) in~\cite{BenderOrszag}. We find
\be
B_n=\left(-\frac{2}{3}\right)^{n-1}(1+n)\left(\frac{3}{2}-H_{1+n}\right) \;,
\ee
which is very similar to the LCF case. Summing all orders we find
\be\label{ABmatrix}
\mathbb{P}=\frac{1}{2}{\bf N}_0\cdot\begin{pmatrix}1&B\\0&1+A\end{pmatrix}\cdot{\bf N}_f \;,
\ee
where
\be
A=-\frac{b_0^2a_0^2\mathcal{T}}{1+\frac{2}{3}b_0a_0^2\mathcal{T}}\mathcal{I}
\ee
and
\be
\begin{split}
B=\frac{-b_0^2}{\left(1+\frac{2}{3}b_0a_0^2\mathcal{T}\right)^2}&\bigg(b_0a_0^2\mathcal{T}\left[1+\frac{1}{3}b_0a_0^2\mathcal{T}\right] \\
&-\frac{3}{2}\ln\left[1+\frac{2}{3}b_0a_0^2\mathcal{T}\right]\bigg)\mathcal{I} \;,
\end{split}
\ee
where
\be
\mathcal{I}=\frac{1}{a_0^2}\int_0^\infty\ud\gamma\;\gamma^2\mathcal{J}_0(\gamma) \;.
\ee
For small $a_0$ we have $\mathcal{I}\approx2/5$, and $\mathcal{I}\approx35a_0/(24\sqrt{3})$ for large $a_0$. As a check, note that by summing over the final spin we have $\mathbb{P}={\bf N}_0\cdot\{1,0\}=1$, which also explains why there are only two nontrivial elements. For an unpolarized initial electron we have $\mathbb{P}=(1/2)\{1,B\}\cdot{\bf N}_f$, so $B$ gives an induced polarization. Hence, without RR the degree of polarization can only be reduced, but with RR we find a nonzero induced polarization.
For a very long pulse we find
\be
A\approx-\frac{3}{2}b_0\mathcal{I} \qquad B\approx-\frac{3}{4}b_0^2\mathcal{I} \;.
\ee 
This should be compared with the Sokolov-Ternov effect. For low-energy electrons, $b_0\ll1$, the induced polarization is a small effect. The reason for this is that RR leads to a reduction of the (lightfront) longitudinal momentum.

For larger $b_0$ we can use the resummation or the integro-differential approach. We have used the Borel-Leroy method for resumming the $b_0$ expansion. For $B$ we have~\eqref{BorelLoreyTransform} with $b_n=4+(3/2)n$ for $n\geq2$, while for $A$ we have $b_n=3+(3/2)n$ for $n\geq1$. For both terms $c=1.588...$ and $[40/(40+\text{ceil.}[b_n]+1)]$.

The results are shown in Fig.~\ref{STRRcircFig}. At $b_0=0.3$ the induced polarization is on a $\sim1\%$ level. This is not a large effect compared to the constant-field case, but it is non-negligible, and it is a clear signal of RR since without RR there would be no induced polarization.

\subsection{Circular perturbative}

\begin{figure}
\includegraphics[width=\linewidth]{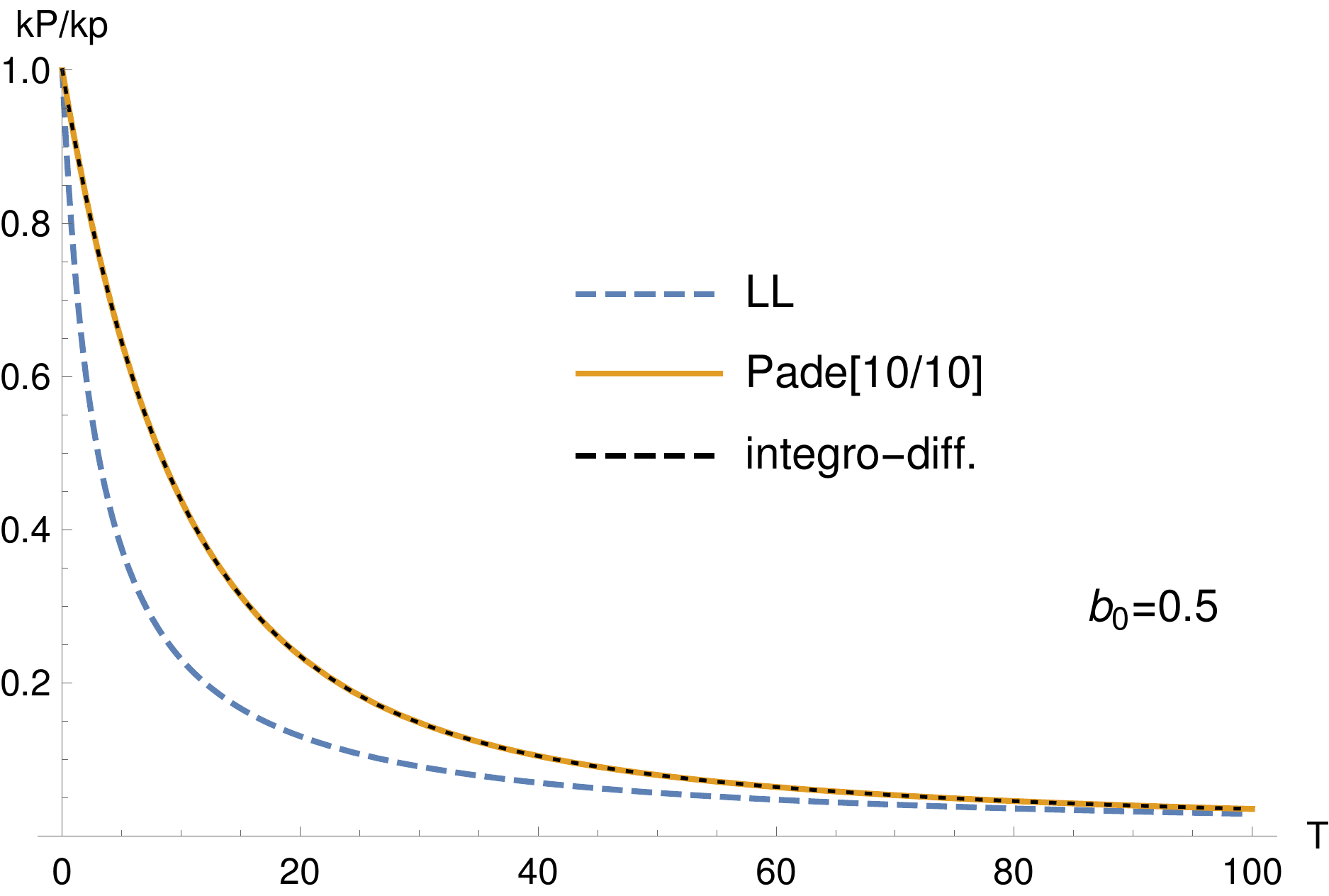}
\includegraphics[width=\linewidth]{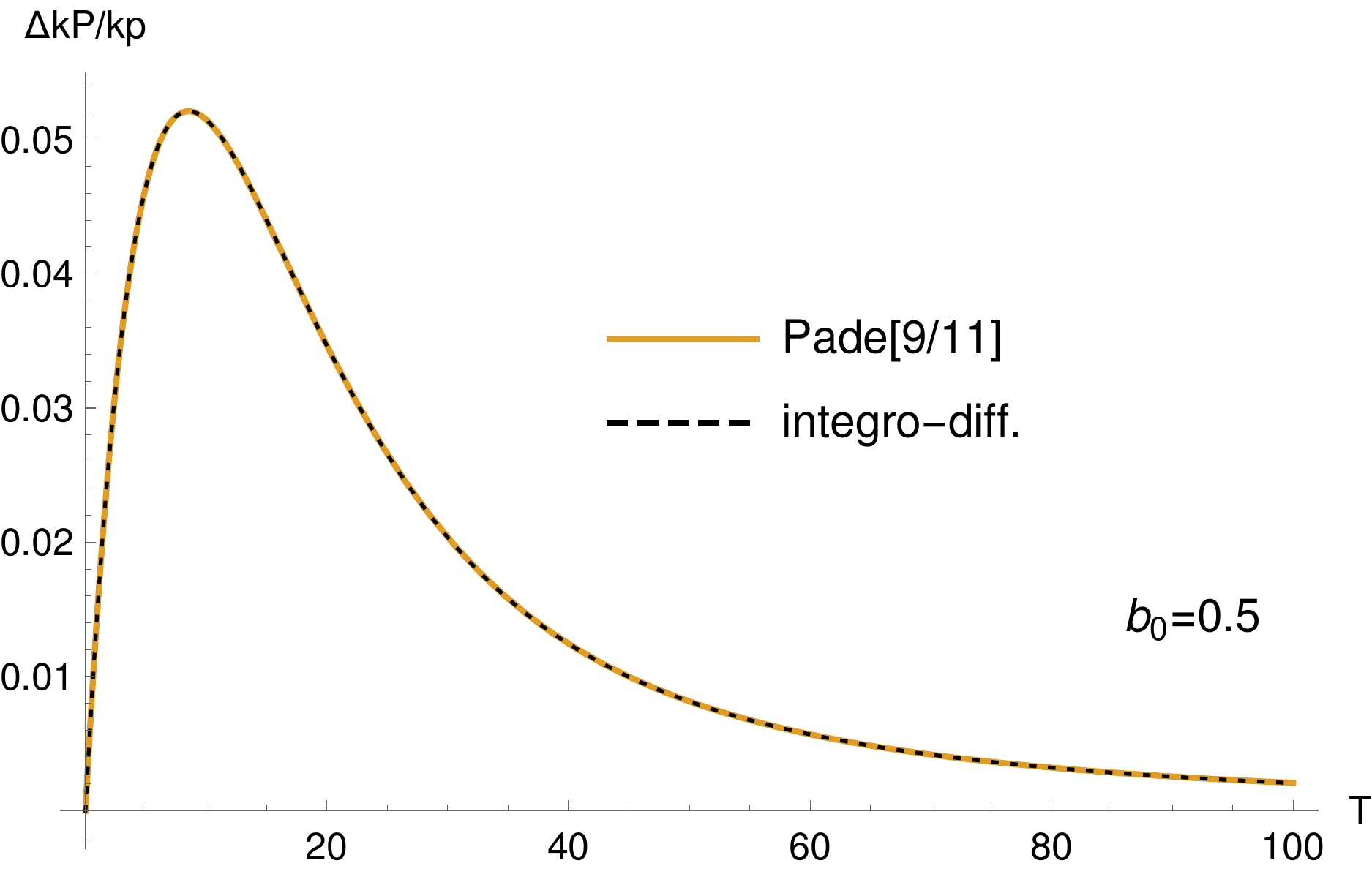}
\caption{Momentum expectation value to leading order in $a_0\ll1$.}
\label{per05fig}
\end{figure}

In this section we will expand each order in $\alpha$ to leading order in $a_0\ll1$. In this regime the $\mathcal{J}_i$ functions become very simple,
\be
\mathcal{J}_0=a_0^2\frac{\gamma}{4}(2-\gamma)
\qquad
\mathcal{J}_1=\frac{a_0^2}{2}
\qquad
\mathcal{J}_2=a_0^2\frac{1-\gamma}{2}
\ee
and $\mathcal{J}_i=0$ for $\gamma>2$. Since ${\bf M}^{\rm C}$ and ${\bf M}^{\rm L}$ are proportional to $a_0^2$ it is natural to define the following effective expansion parameter
\be\label{TcircPer}
\text{circular field, perturbative:}\qquad
T=\alpha a_0^2\Delta\phi \;.
\ee
Since $a_0^2\ll1$, the pulse length would have to be much longer in order to have a not small $T$. This makes perhaps this regime less interesting from an experimental point of view. It can also be challenging for calculations, because small $a_0$ in general makes the corrections to the incoherent product approximation more important, so the pulse length has to be extra long to compensate for this. However, there is an interesting difference compared to the other cases: The $b_0$ expansions are no longer asymptotic, but have a finite radius of convergence. We can therefore resum both the $b_0$ and the $\alpha$ expansions with Pad\'e approximants. In Fig.~\ref{per05fig} we have resummed the $b_0$ expansion with $[46/52]$ and $[46/51]$ for $\langle kP\rangle(\uparrow)+\langle kP\rangle(\downarrow)$ and $\langle kP\rangle(\uparrow)-\langle kP\rangle(\downarrow)$, respectively.

\section{Resummation of LAD}\label{Resummation of LAD}

\begin{figure*}
\includegraphics[width=\linewidth]{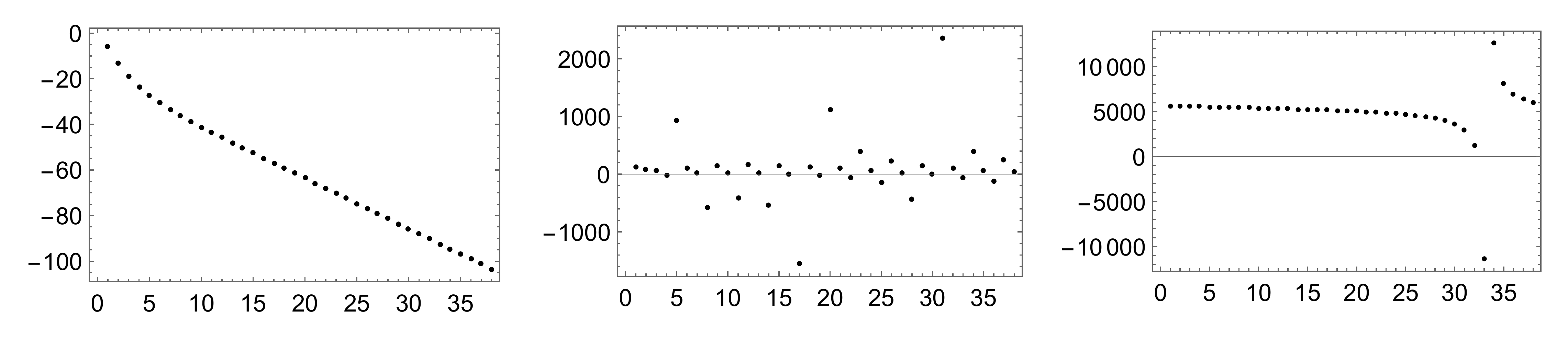}
\caption{Ratios of neighboring odd-order coefficients for $g_0$, $g_0^{(2[n+1]+1)}/g_0^{(2n+1)}$, with $u=\{1/10,11,75\}$ from left to right. The corresponding plots for the even orders, $g_0^{(2[n+1])}/g_0^{(2n)}$, have a similar behavior.}
\label{uResumRatiosFig}
\end{figure*}

\begin{figure}
\includegraphics[width=\linewidth]{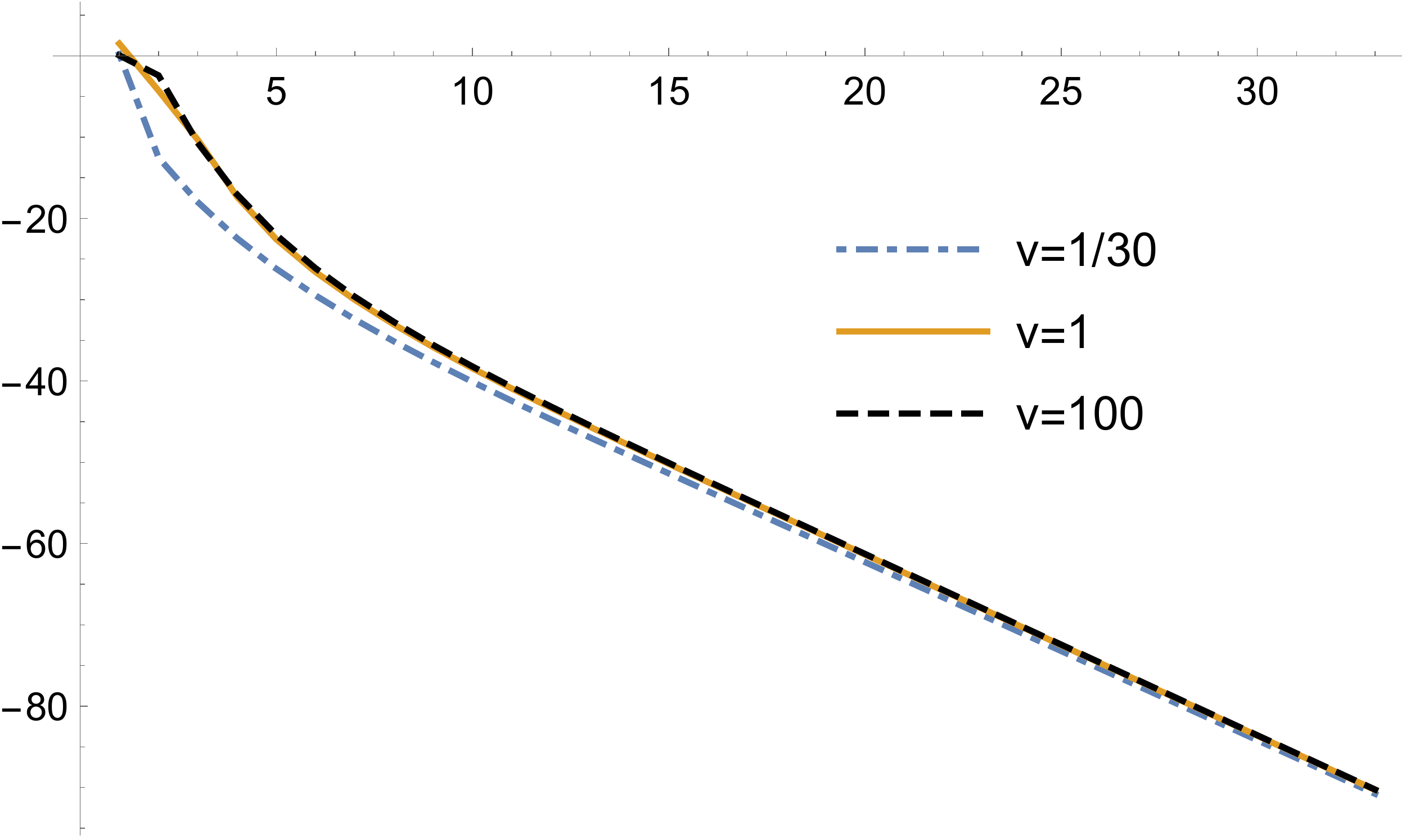}
\caption{Ratios of neighboring coefficients for $h_0(v)$, $h_0^{(n+1)}(v)/h_0^{(n)}(v)$.}
\label{vResumRatiosFig}
\end{figure}

\begin{figure}
\includegraphics[width=\linewidth]{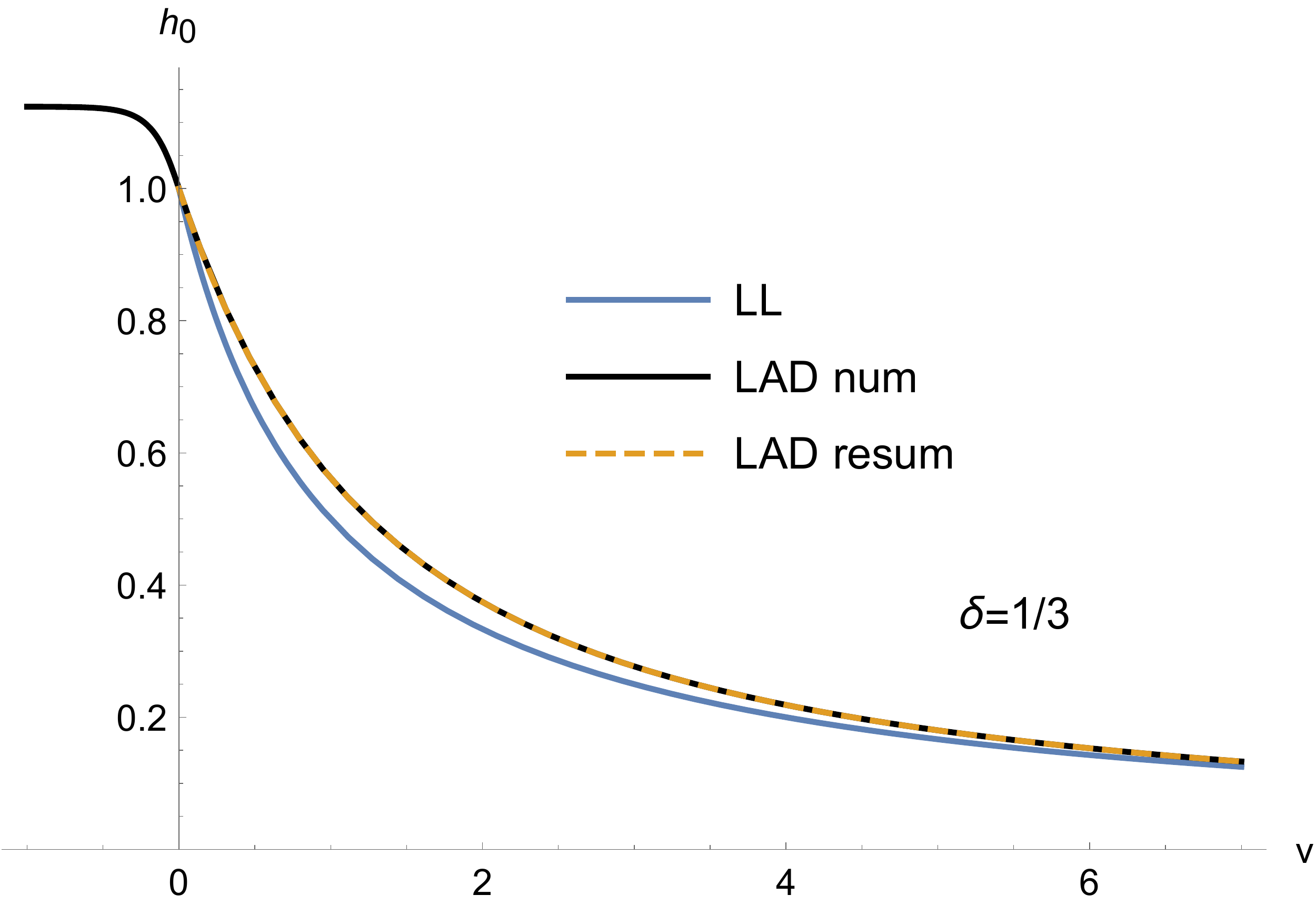}
\includegraphics[width=\linewidth]{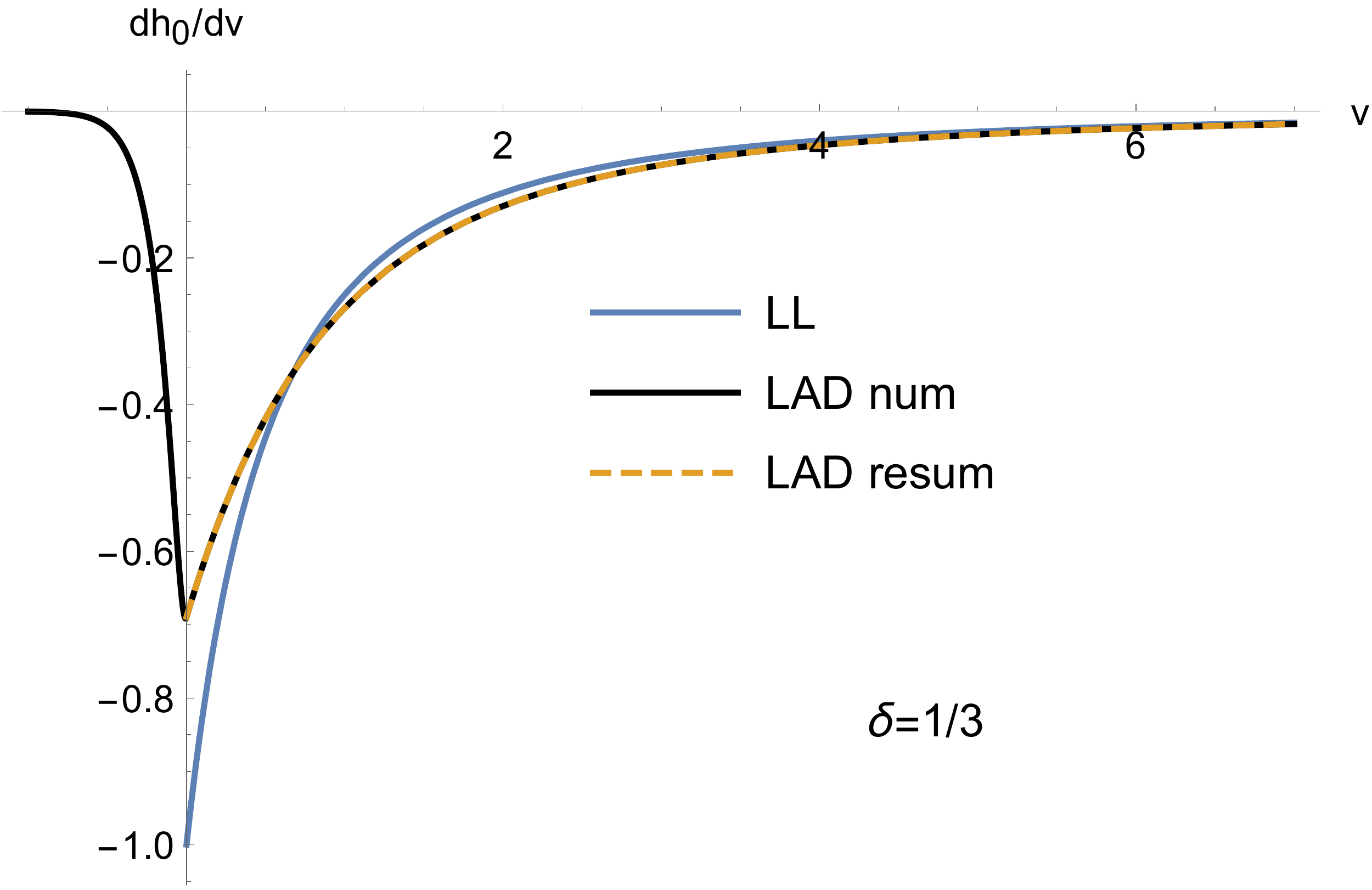}
\caption{Resummation of LAD using Borel-Pad\'e$[15/15]$. The numerical results have been obtained by integrating LAD backwards with the resummation result at $v=15$ as final conditions.}
\label{h0plotFig}
\end{figure}

\begin{figure}
\includegraphics[width=\linewidth]{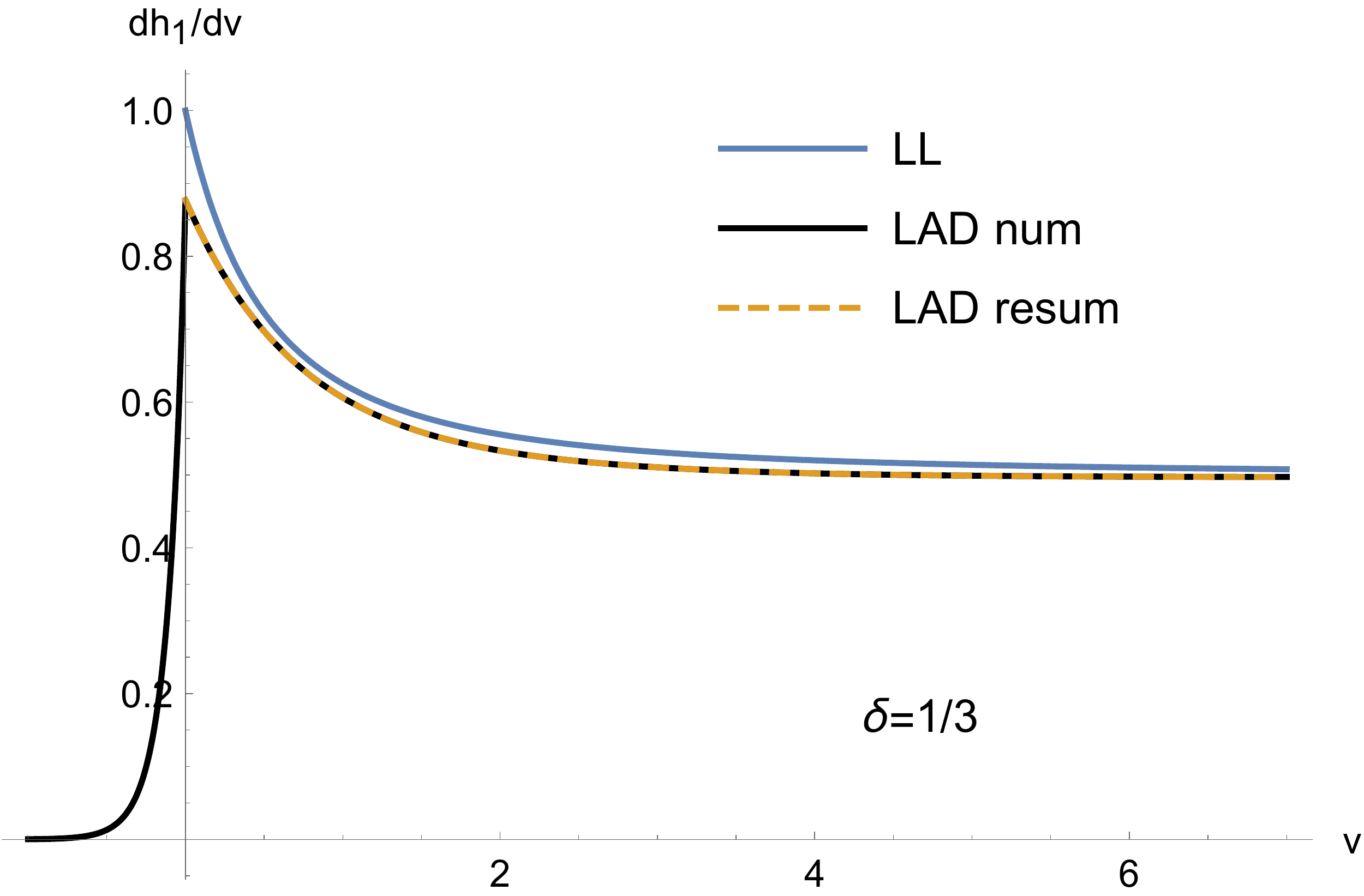}
\caption{As Fig.~\ref{h0plotFig} for second term.}
\label{h1plotFig}
\end{figure}

Both the classical LL and our quantum approximation give series in $\alpha$ which can be resummed with Pad\'e approximants. However, we know/suspect that the full QED expansion is asymptotic. If this is the case then LL cannot be the exact equation, but LAD might be since it leads to asymptotic series in $\alpha$. In~\cite{Zhang:2013ria} it was shown how the nonrelativistic LAD can be resummed with the Borel method. In this section we will consider the full LAD expansion. Since we cannot obtain a compact/closed form expression for each order, we will restrict to a constant-crossed field for simplicity, $f_{\mu\nu}=k_\mu a'_\nu-a'_\mu k_\nu$ with $a'$ constant. 

LAD is given by
\be
\ddot{x}^\mu=f^{\mu\nu}\dot{x}_\nu+\frac{2}{3}\alpha(\dddot{x}^\mu+\ddot{x}^2\dot{x}^\mu) \;.
\ee 
Since we consider a constant field, only derivatives of the position appear. We denote $P^\mu=\dot{x}$. We write the solution as
\be
P^\mu(\tau)=\left(g_0\delta^\mu_\nu+g_1\frac{{f^\mu}_\nu}{\chi_0}+g_2\frac{{(f^2)^\mu}_\nu}{\chi_0^2}\right) P^{(0)\nu} \;,
\ee
where $P^{(0)}$ is the momentum at $\tau=0$, $\chi_0=\sqrt{P^{(0)}f^2P^{(0)}}$ and $g_i(\tau)$ are functions to be determined. From the on-shell condition $P^2=1$ and $f^3=0$ we find
\be
g_2=\frac{1+g_1^2-g_0^2}{2g_0} \;.
\ee
Instead of proper time $\tau$, it is natural to use a rescaled lightfront time 
\be
u(\tau)=\frac{\chi_0}{kP^{(0)}}kx(\tau) \qquad \frac{\ud}{\ud\tau}=\chi_0 g_0\frac{\ud}{\ud u} \;.
\ee
We also define
\be
\delta=\frac{2}{3}\alpha\chi_0 \;.
\ee
Projecting LAD with $P^{(0)}f^2$ and $P^{(0)}f$ gives us two equations for the two unknown functions
\be\label{g01eq}
\begin{split}
g_0'(u)&=\delta(\partial_u[g_0g_0']+P'^2g_0^2) \\
g_1'(u)&=1+\delta(\partial_u[g_0g_1']+P'^2g_0g_1) \;,
\end{split}
\ee
where $F'=\ud F/\ud u$ and 
\be
P'^2=g_0'^2+2g_0'g_2'-g_1'^2 \;.
\ee

We will first attempt to solve~\eqref{g01eq} by a direct expansion in $\delta$. The zeroth order is given by
\be
g_0=1+\mathcal{O}(\delta) \qquad g_1=u+\mathcal{O}(\delta) \;.
\ee
Substituting $g_i=\sum_{n=0}^N\delta^n g_i^{(n)}(u)+\mathcal{O}(\delta^{N+1})$ into the right-hand sides of~\eqref{g01eq} and integrating over $u$ gives $g_i^{(N+1)}(u)$. For the first couple of orders we find
\be\label{g0uExp}
\begin{split}
g_0&=1-u \delta +u^2 \delta ^2+\left(6 u-u^3\right)\delta^3+\left(-18 u^2+u^4\right)\delta^4 \\&+\left(-80 u+36 u^3-u^5\right)\delta^5+\left(392 u^2-60 u^4+u^6\right)\delta^6 \\
&+\left(1520 u-\frac{3524 u^3}{3}+90u^5-u^7\right)\delta^7+\mathcal{O}\left(\delta^8\right)
\end{split}
\ee
\be\label{g1uExp}
\begin{split}
g_1&=u-\frac{u^2 \delta }{2}+\left(-2u+\frac{u^3}{2}\right) \delta ^2+\left(6
   u^2-\frac{u^4}{2}\right) \delta ^3 \\
   &+\left(20
   u-\frac{41 u^3}{3}+\frac{u^5}{2}\right)
   \delta ^4+\mathcal{O}\left(\delta ^5\right) \;.
\end{split}
\ee
Each order is a polynomial in $u$ and the first tens of orders can be obtained quickly. However, the number of terms increases as we go to higher orders and eventually the calculation becomes very slow. We stopped at $\mathcal{O}(\delta^{77})$. To study the (non)-convergence properties of a series one can usually plot the ratios of neighboring coefficients. However, here it turns out to be easier to study the even and odd orders separately. In Fig.~\ref{uResumRatiosFig} we plot the ratios of the odd-order coefficients for a small, moderately large and large value of $u$. For small $u$ we see a typical line for an asymptotic series with factorially growing coefficients with alternating sign. Such a series can be dealt with using the Borel-Pad\'e method. For the moderately large $u$ we have a very different plot with no discernible pattern up to the maximum order calculated. In the large $u$
plot we again see a simple pattern, but this time in the first $\sim20$ orders rather than as an asymptotic limit. This almost constant line is what one expects from a function that is close to LL. We can understand this from~\eqref{g0uExp}: After a long time in the field, $u\gg1$, one may expect the highest power in $u$ at each order in $\delta$ to give a good approximation, i.e.
\be\label{g0LL}
g_0\to\sum_{n=0}^\infty(-\delta u)^n=\frac{1}{1+\delta u}=g_0^{\rm LL} \;.
\ee
Hence, for a sufficiently long pulse we expect the solution to LAD to converge to the LL solution, where the in general asymptotic series can be approximated by a convergent series.
For smaller $u$, on the other hand, we have checked that the $\delta$ expansion can be resummed with the Borel-Pad\'e method. However, as $u$ increases this resummation seems to eventually break down. One might try to find different resummations or a way to obtain higher orders. However, we have used a different approach instead, which seems to work for any $u$.

In this approach we begin by noting that, e.g. from~\eqref{g0LL}, we expect to have a considerable change in the momentum for $\delta u\sim1$. We therefore change variable to $v=\delta u$ and treat $v$ as independent of $\delta$. We also define 
\be
h_0(v)=g_0(u=v/\delta) \qquad h_1(v)=\delta g_1(u=v/\delta) \;.
\ee
We expand the solution as
\be
h_i(v)=\sum_{n=0}^\infty h_i^{(n)}(v)\delta^{2n} \;.
\ee
Note that this is not the same as repeatedly substituting LAD into itself to remove higher-order derivatives, an approach that has recently been studied in~\cite{Ekman:2021vwg}.
To zeroth order in $\delta$, the solution is now given by LL
\be
h_0=\frac{1}{1+v}+\mathcal{O}(\delta^2)=h_0^{\rm LL}+\mathcal{O}(\delta^2)
\ee
and
\be
h_1=v-\frac{v^2}{2(1+v)}+\mathcal{O}(\delta^2)=h_1^{\rm LL}+\mathcal{O}(\delta^2) \;.
\ee
In this approach, each order is not just a polynomial in $v$. Fortunately, though, we can still find simple explicit expressions for each order. The first couple of orders are given by
\be
h_0=\frac{1}{v+1}+\frac{6 v}{(v+1)^3}\delta^2-\frac{4 v \left(60+6 v+11 v^2\right)}{3
   (1+v)^5}\delta^4+\mathcal{O}(\delta^6)
\ee
and
\begin{widetext}
\be
h_1=\frac{v (2+v)}{2 (1+v)}+\left(\frac{4 \log
   (1+v)}{1+v}-\frac{3 v \left(2+2
   v+v^2\right)}{(1+v)^3}\right) \delta ^2+2 v
   \left(\frac{12 \log
   (1+v)}{(1+v)^3}+\frac{30-48 v+4 v^2+28
   v^3+11 v^4}{3 (1+v)^5}\right) \delta ^4+\mathcal{O}(\delta^6)
\ee
\end{widetext}
Note that using $v$ instead of $u$ gives series in $\delta^2$ rather than $\delta$. We have obtained the first $34$ terms, i.e. up to $\mathcal{O}(\delta^{66})$. The expressions become very long, we just note that in the limit $v\to\infty$ all higher orders scale as $1/v$ compared to the zeroth (LL) order. If we go back to the original variable $u$, then each function $h_i^{(n)}(v=\delta u)$ provides a partial resummation of the $\delta$ expansion from the first approach. In particular, the exact solution to LL is already included. Consequently, the ratios of neighboring coefficients, $h_0^{(n+1)}(v)/h_0^{(n)}(v)$ shown in Fig.~\ref{vResumRatiosFig}, now have a simple behavior (typical for alternating and factorially divergent series) for both small and large $v$, in contrast to Fig.~\ref{uResumRatiosFig}.

This means that for an arbitrary value of $v$ we can resum the $\delta$ expansion using the Borel-Pad\'e method. We have used $[15/15]$ approximants. The results are shown in Fig.~\ref{h0plotFig} and~\ref{h1plotFig} for $\delta=1/3$. The reason for choosing this rather large value of $\delta$ is to be able to clearly distinguish LAD and LL. For smaller $\delta$ it is difficult to see the difference because it is given by a function of $\delta^2$. After a very long time in the field, $v\gg1$ or $u\gg1/\delta$, the result converges to LL. We have also compared the resummation with a numerical solution to LAD. Since LAD is numerically unstable when integrated forwards in time, we have integrated LAD backwards in time~\cite{Hartemann:1996zza,KogaBackwardsIntegration} with the resummed result $h_i^{\rm resum}(v_f\gg1)$ at one instant of time $v_f$ as final conditions. As shown in Fig.~\ref{h0plotFig} and~\ref{h1plotFig}, we find excellent agreement between the numerical solution and the resummation. The considered value of $\delta$ is quite large, but since the resummation is based on a $\delta\ll1$ expansion, it would be even more precise for smaller $\delta$. For the numerical solution we have continued passed $v<0$ assuming that the field is zero for $v<0$, to illustrate the fact that the solution has pre-acceleration. This means that $P_\mu^{(0)}$ is strictly speaking not the (exact) initial momentum even if the field is zero at $u<0$.     
Thus, the $\delta$ series gives a pre-acceleration solution without runaway. To obtain the runaway solutions one would start with e.g. a trans-series-like ansatz; see e.g. the recent papers~\cite{Borinsky:2021hnd,Du:2021fok} for trans-series solutions to differential equations.

\section{Conclusions}

We have studied RR using our recently proposed resummation methods. In~\cite{Torgrimsson:2021wcj} we studied the electron momentum expectation value for a constant-crossed field and with the initial electron either unpolarized or polarized along the magnetic field. Here we have considered the momentum for a circularly polarized monochromatic field with $a_0=1$ and the initial electron unpolarized or polarized along the laser propagation direction. We have found that, as a function of an appropriate effective pulse length parameter (proportional to $\alpha$), both the spin-averaged momentum $P_\LCm(\uparrow)+P_\LCm(\downarrow)$ and the difference $P_\LCm(\uparrow)-P_\LCm(\downarrow)$ look similar, even though both the field and spin direction are different.

In both cases, we have also studied the polarization of the final electron. For a constant field and without RR, the polarization is given by the Sokolov-Ternov result, which is $\approx0.92$ asymptotically. Here we find that with RR there is still a nonzero degree of polarization asymptotically, but this asymptotic value is reached on a much shorter time scale and is significantly lower.
For a circularly polarized monochromatic field and without RR, there is no induced polarization to leading order~\cite{KotkinUseLoops,Kotkin:1997me,Torgrimsson:2020gws}, because the loop cancels the contribution from photon emission. However, here we find that with RR there is a nonzero degree of polarization. 

Finally, in order to gain some insights into what one might find if one could find a way to include the terms that have been neglected in the incoherent product approach, we have shown how to use resummation methods for LAD.

\acknowledgements
On the day that this manuscript was submitted, \cite{Ekman:2021eqc} appeared on the arXiv. The subject of~\cite{Ekman:2021eqc} is very similar to our section about resummation of LAD. Our Fig.~\ref{h0plotFig} is very similar to their Fig.~7 (even the plot style happens to be similar). The main difference is how LAD has been resummed. \cite{Ekman:2021eqc} resummed terms appearing in the differential equation, while here we have resummed the solution directly. We have checked, though, that our expansions for $g_0'(u=0)$ and $g_1'(u=0)$ agree with Eq.~(7) in~\cite{Ekman:2021eqc}. I thank Robin Ekman and Anton Ilderton for discussions about this.

\end{document}